\documentclass{cernyrep}
\usepackage{texnames}
\usepackage[T1]{fontenc}
\usepackage[bookmarks, colorlinks=true, linktoc=page, pdftex, linkcolor=black, citecolor=black, urlcolor=blue]{hyperref}
\usepackage{url}
\usepackage{hyperref}
\usepackage{fancyhdr}
\fancyhfoffset{4 mm}
\fancypagestyle{ARTTITLE}{%
\fancyhf{} 
\lhead{\small{CERN Accelerator School Proceedings ---
{\it Advanced Accelerator Physics} ---  Spa,  Belgium, 2024}}
\lfoot{\hspace{3mm} Available online at \url{https://cas.web.cern.ch/previous-schools}}
\rfoot{\thepage\hspace*{3mm}}
 
}

\usepackage{varwidth}
\usepackage{subfig}
\usepackage{xcolor}
\usepackage{hyperref}
\usepackage{amsmath}
\usepackage[sort&compress,numbers]{natbib}

\usepackage{listings}
\usepackage[labelfont=bf]{caption}
\usepackage{varwidth}
\usepackage{xcolor}
\begin{document}
\title{Beam-beam}

\author{X. Buffat}

\institute{CERN, Geneva, Switzerland}

\maketitle 

\begin{abstract}
The interaction of the two beams in a collider leads to a variety of effects that may limit the performance of the machine. This lecture introduces the basic aspects necessary to understand the design of modern colliders.
\end{abstract}
\maketitle
\thispagestyle{ARTTITLE}

\section{Introduction}
\begin{figure}
\begin{center}
    \subfloat[electron-positron]{\includegraphics[width=0.45\linewidth]{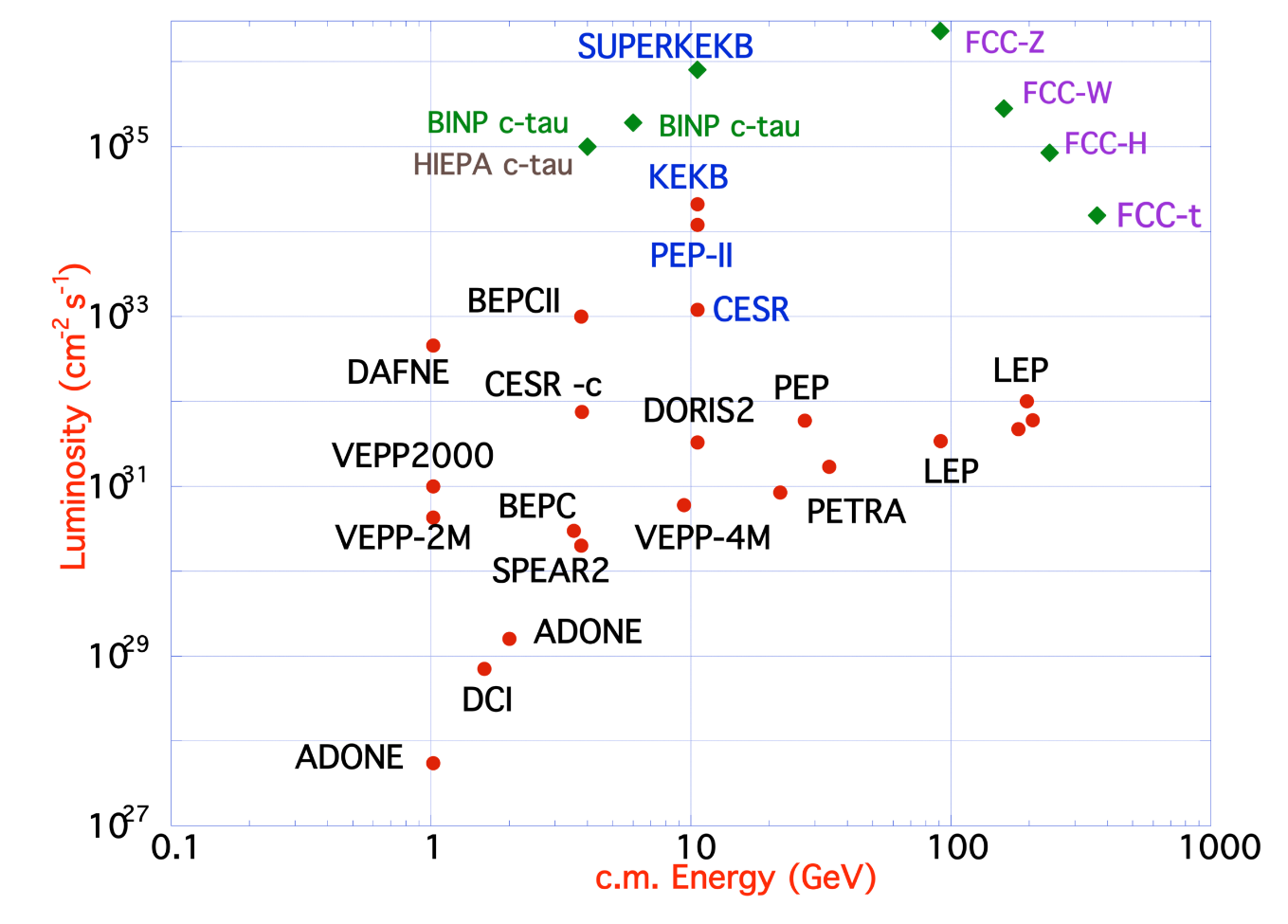}\label{fig-circular-ep}}
    \qquad
    \subfloat[hadron]{\includegraphics[width=0.45\linewidth]{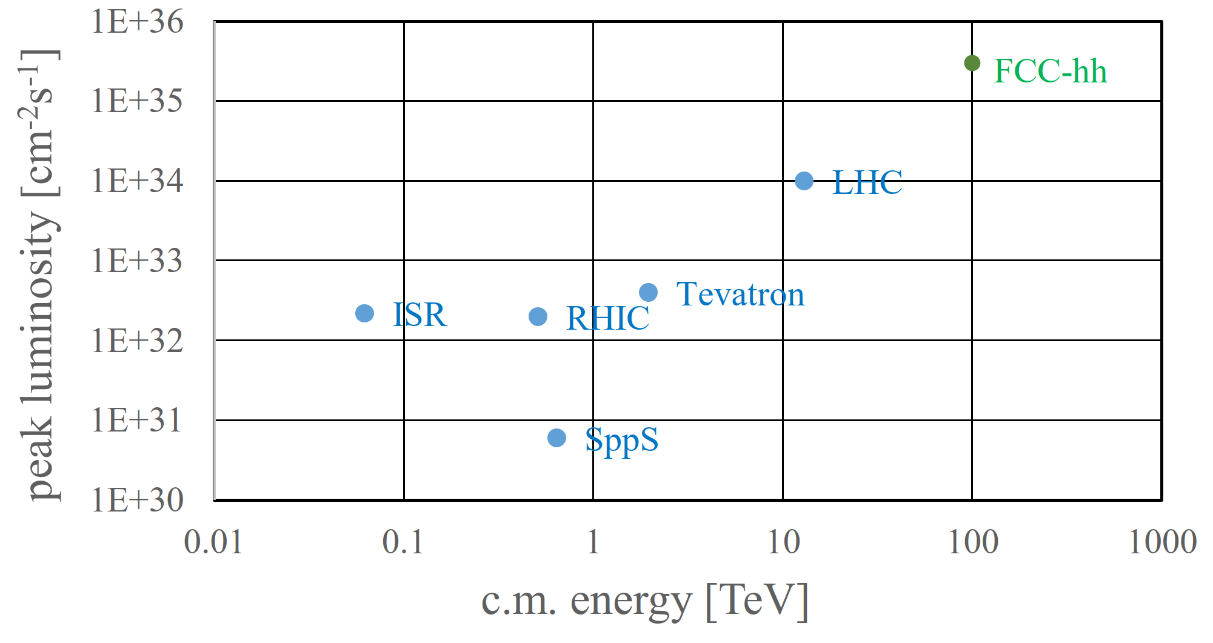}\label{fig-circular-pp}}
    \qquad
\end{center}
\caption{List of circular colliders \cite{FCCBenedikt,wikilist}}
\end{figure}
Particle colliders are without doubt some of the most important tools to study particle physics. They allow to generate collisions with the highest center of mass energy in copious amounts and in controlled conditions, such that high energy physics processes may be studied in details. The richness of the field is well illustrated by the amount of machines built since the first colliders in the 60s. The large majority are circular electron-positron colliders (Fig.~\ref{fig-circular-ep}). They are usually preferred as the initial state of the collision is known with a high precision. Circular hadron colliders are the second most popular (Fig.~\ref{fig-circular-pp}), as their energy reach goes beyond the one of electron-positron colliders, at the cost of more complicated initial states since hadrons are composite particles. Hadron colliders are also built to study processes specific to hadrons, such as nuclear physics or quark-gluon plasmas. For such studies, electron-hadron colliders are also particularly suited. Only one collider of this type existed, the Hadron-Elektron-RingAnlage (HERA)~\cite{HERA}. Another one is planed to operate in a near future: the Electron-Ion Collider (EIC)~\cite{EIC}. A~single linear electron-positron collider was build, the Standford Linear Collider (SLC)~\cite{SLC}. Note though that it was not strictly linear, as it featured a single linear accelerator and two arms bringing the two beams against each other. Linear electron-positron colliders have the advantage over circular colliders that they do not suffer from the energy loss due to synchrotron radiation, at the cost that each accelerated particle gets only a single chance to collide with the other beam.

With respect to other particle accelerators or storage rings, colliders feature a key difference: The~beam-beam force. This force is non-linear, dynamic and usually strong, thus leading to lifetime degradation, beam blowup and even emission of radiations. Colliders are usually build specifically to maximise the~amount of collisions (i.e. the luminosity) while maintaining the detrimental effects of beam-beam interactions under control. The main performance limitations arise, directly or indirectly, from the beam-beam force. Its understanding was therefore key to drive the design of present colliders and remains crucial to push their performance and design even more performing ones in the future.

In the next chapter, we'll describe the main models for the beam-beam force. We then discuss its first order impact on the beam orbit and optics as well as specific issues linked to the dynamic nature of the force. Finally, we discuss non-linear effects, radiations and mitigation, which leads us to understanding of the design of modern colliders.
\section{The beam-beam force}
\begin{figure}
\begin{center}
\includegraphics[width=0.6\linewidth]{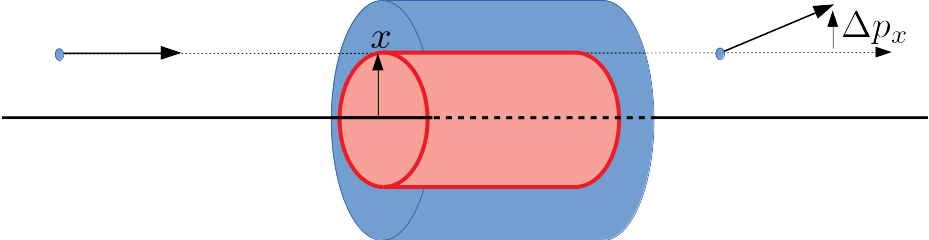}
\end{center}
\caption{Illustration for the computation of the force experienced by a particle with position $x$ with respect to the~center of a uniformly charge cylinder (blue). The co-axial cylinder considered to apply Gauss law is drawn in red.}
\label{fig-cylinder}
\end{figure}
\begin{figure}
\begin{center}
    \subfloat[Cylinder]{\includegraphics[width=0.45\linewidth]{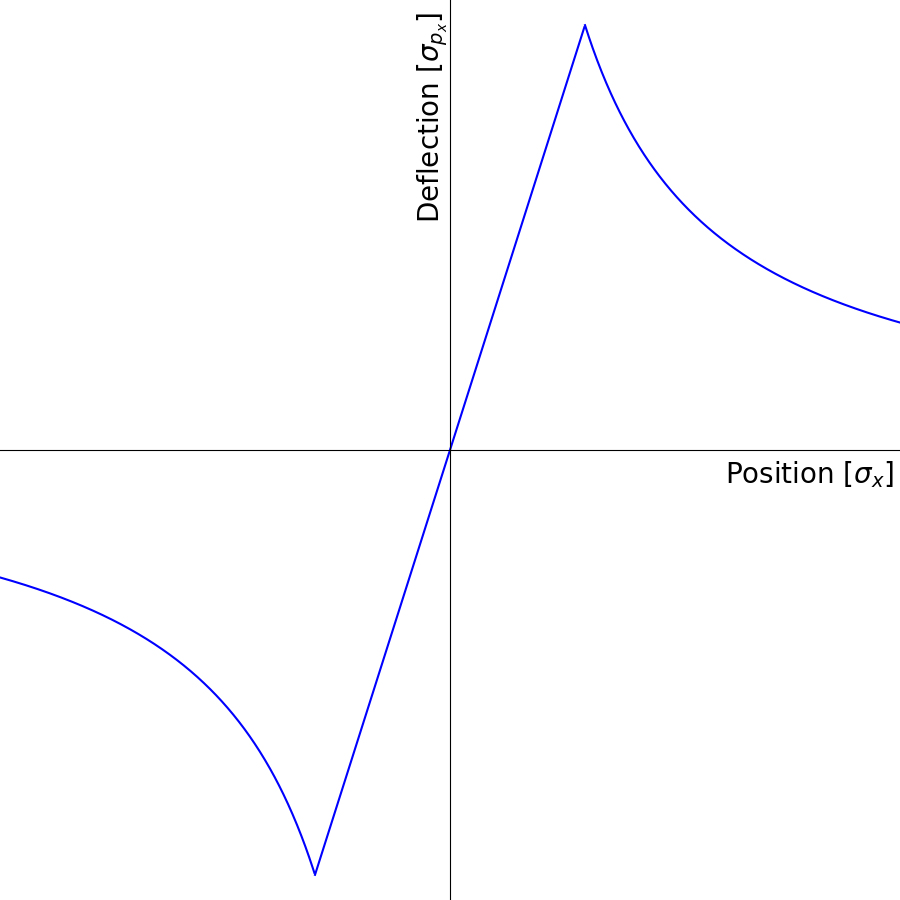}\label{fig-force-cylinder}}
    \qquad
    \subfloat[Gaussian]{\includegraphics[width=0.45\linewidth]{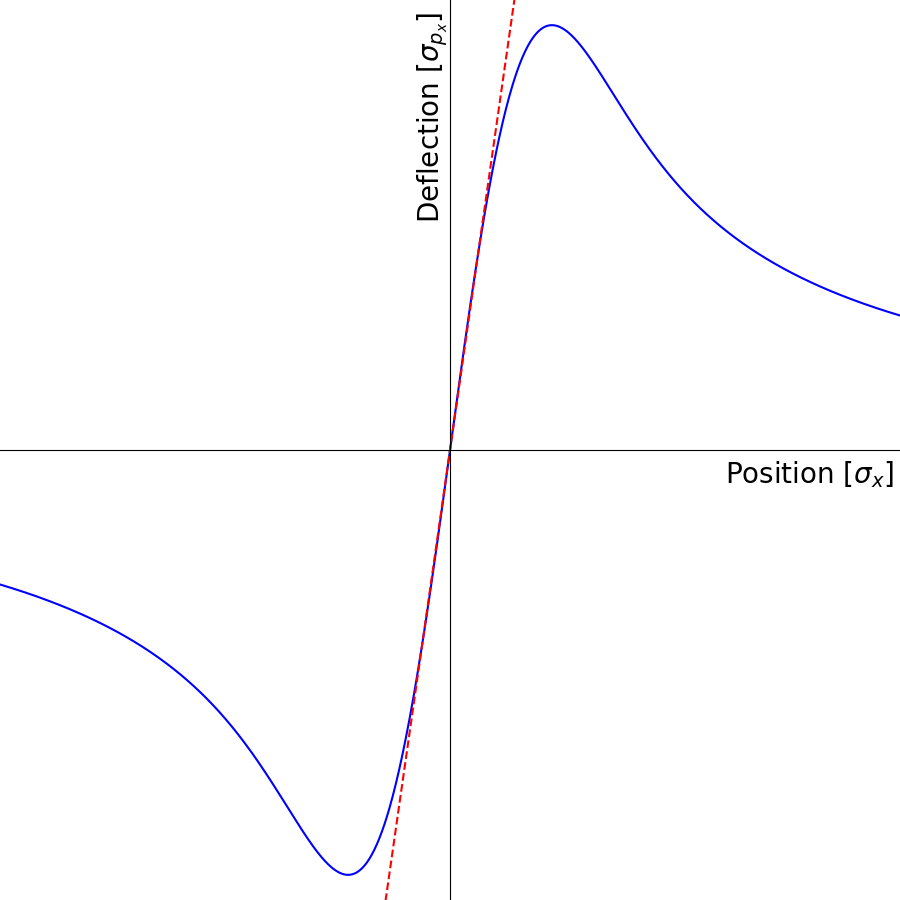}\label{fig-force-gauss}}
    \qquad
    \subfloat[Gaussian $\color{blue}\sigma_x\color{black} = 100\cdot\color{red}\sigma_y$ ($\beta_x = 50\cdot\beta_y$, $\epsilon_x = 200\cdot\epsilon_y$)]{\includegraphics[width=0.45\linewidth]{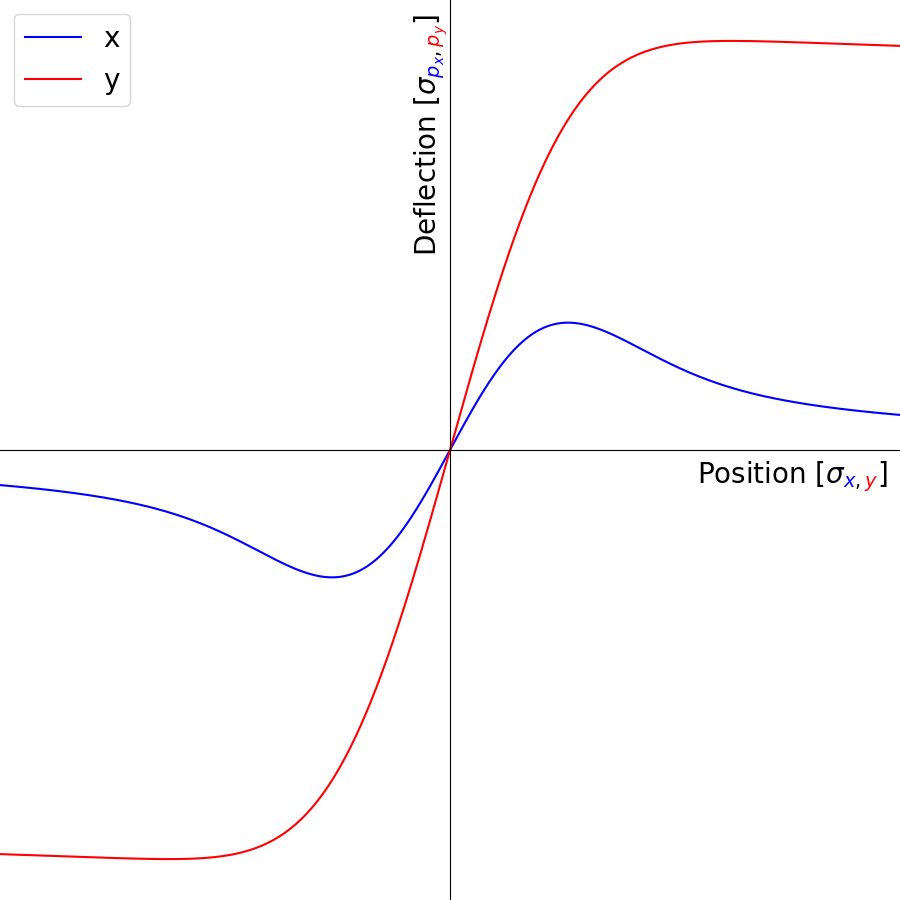}\label{fig-force-flat}}
    \qquad
    \subfloat[Long-range]{\includegraphics[width=0.45\linewidth]{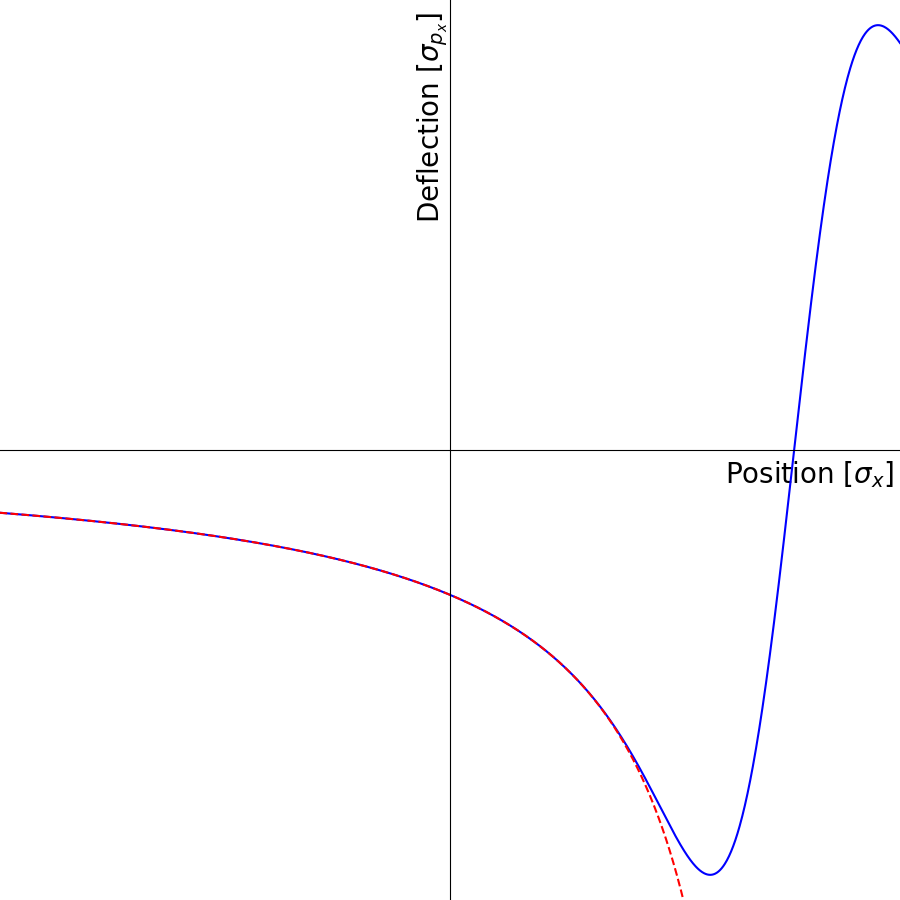}\label{fig-force-LR}}
    \qquad
\end{center}
\caption{Illustration of the beam-beam force for different configurations. The linear and $1/r$ approximations are marked in dashed red in Figs.~\ref{fig-force-gauss} and~\ref{fig-force-LR} respectively.} \label{sec-bbforce}
\label{fig-force}
\end{figure}
To understand the beam-beam force we start by computing the force experienced by a charged particle going through an infinitely long and uniformly charged cylinder (Fig.~\ref{fig-cylinder}). Using the integral form of Gauss law on a co-axial cylinder with a radius corresponding to the position of the particle, we easily find that the force depends linearly on the position of the particle with respect to the center if it is inside the cylinder and decaying as $1/x$ when outside, as illustrated in Fig.~\ref{fig-force-cylinder}. Similarly, one finds a comparable but smoother behaviour when considering a cylinder with a Gaussian distribution of charges (Fig.~\ref{fig-force-gauss})~\cite{bbforceround}. Using Newton's law we write the deflection that corresponds to the beam-beam force:
\begin{equation}
\Delta p_x = \pm\frac{2Nr_0}{\gamma}\frac{x}{x^2+y^2}\left( 1-e^{-\frac{x^2+y^2}{2\sigma^2}}\right),
\end{equation}
with $r_0$ the classical radius, $\gamma$ the relativistic factor, $N$ the number of charges in the opposing beam, $\sigma$ the r.m.s. size of the other beam and $x$ and $y$ the transverse coordinate with respect to the center of the~opposing beam's distribution. Clearly, $x$ and $y$ can be exchanged to obtain the force in both transverse planes. The force is attractive (i.e. negative sign) if the charge of the two beams have opposite signs, it is repulsive (i.e positive sign) otherwise. It is convenient to characterise the strength of the beam-beam force by the slope of the force at the center, as illustrated in Fig.~\ref{fig-force-gauss}. We have:
\begin{equation}
\Delta p_x = \pm k_{bb}\cdot x,
\end{equation}
where we have introduced to beam-beam strength analogously to the one of a quadrupole: 
\begin{equation}
k_{BB} = \pm\frac{r_0N}{\gamma\sigma^2} \label{eq-kbb}.
\end{equation}
Using the formula for a quadrupole error~\cite{accHandbook}, we can then write the induced tune shift $\Delta Q_{BB}$:
\begin{equation}
\cos(2\pi(Q_0 + \Delta Q_{BB})) = \cos(2\pi Q_0) - \beta_0^*k_{BB}\sin(2\pi Q_0), \label{eq-bbtuneshift}
\end{equation}
with $Q_0$ the unperturbed tune. For small enough strength, one may write
\begin{equation}
\Delta Q_{BB} \approx \mp \frac{Nr_0}{4\pi\epsilon_n}\equiv \xi \label{eq-bbparameter},
\end{equation}
which corresponds to the definition of the beam-beam parameter. We note that often the actual beam-beam tune shift is different from the beam-beam parameter, since the approximation of small tune shift used to go from Eq.~\eqref{eq-bbtuneshift} to Eq.~\eqref{eq-bbparameter} is often not valid (see Section~\ref{sec-dynamic}). The beam-beam parameter remains a good way of quantify the strength of the beam-beam force independently of its impact on the beam dynamics.

If the beam size at the interaction point is different in the two planes, the beam-beam forces takes a more complicated form~\cite{bassetierskine}:
\begin{equation}
\Delta p_y +i \Delta p_x = \pm\frac{4Nr_0}{\gamma}\sqrt{\frac{\pi}{2(\sigma_x^2-\sigma_y^2)}}\left(w\left(\frac{x+iy}{\sqrt{2(\sigma_x^2-\sigma_y^2))}}\right)-e^{-\frac{x^2}{2\sigma_x^2}-\frac{y^2}{2\sigma_y^2}}w\left(\frac{\frac{\sigma_y}{\sigma_x}x+i\frac{\sigma_x}{\sigma_y}y}{\sqrt{2(\sigma_x^2-\sigma_y^2)}}\right)\right) \label{eq-bassetierkin}
\end{equation}
where we have introduced the beam sizes in the two transverse planes $\sigma_x$ and $\sigma_y$ as well as the Faddeeva function $w(\cdot)$, a variation of the complex error function~\cite{faddeeva}. The force is illustrated in Fig.~\ref{fig-force-flat}, the beam-beam parameter in each plane become
\begin{equation}
\xi_{x,y} = \mp\frac{Nr_0\beta^*_{x,y}}{2\pi\gamma\sigma_{x,y}(\sigma_x+\sigma_y)},
\end{equation}
which in the frequent case of flat beams ($\sigma_y \ll \sigma_x$), simplifies to
\begin{equation}
\xi_{x} \approx \mp\frac{Nr_0}{2\pi\gamma\epsilon_x}\text{,}~\xi_{y} \approx \mp\frac{Nr_0\beta^*_y}{2\pi\gamma\sigma_y\sigma_x}.
\end{equation}
\subsection{Coherent beam-beam force}
For a number of effects, it is relevant to consider the beam-beam force averaged over the other beam's particle distribution. This is the case for example when considering the orbit change due to a collision with an offset with respect to the beam center. The averaged force is called the coherent beam-beam force and can be obtained with the same formula (Eq.~\eqref{eq-bassetierkin}) but using so-called effective beam sizes~\cite{hirataCoherent}:
\begin{equation}
\sigma_i \longrightarrow \sqrt{\sigma_{i,1}^2+\sigma_{i,2}^2}, \label{eq-effsize}
\end{equation}
where the subscript $i$ refers to the plane $x$ or $y$ and the two beam sizes correspond to the two beams.
\subsection{Long-range interactions}
In colliders featuring multiple bunches and long common chambers (Section~\ref{sec-sep}), the beams may cross in locations where collisions are not wanted, we talk about parasitic encounters. In order to minimise the impact of these parasitic encounters, a transverse separation between the beams is introduced. This separation must be significantly larger than the beam size to avoid collisions and weaken the beam-beam force. In this configuration, the beam-beam force may be approximated by the $1/r$ dependence (Fig.~\ref{fig-force-LR}). Similarly to the head-on case, one may derive formulas to characterise the strength of long-range interactions~\cite{peggs}.
\subsection{Bunch length effects}
\begin{figure}
\begin{center}
    \subfloat[Beam size around the focal point for low (blue) and high (red) $\beta^*$.]{\includegraphics[width=0.35\linewidth]{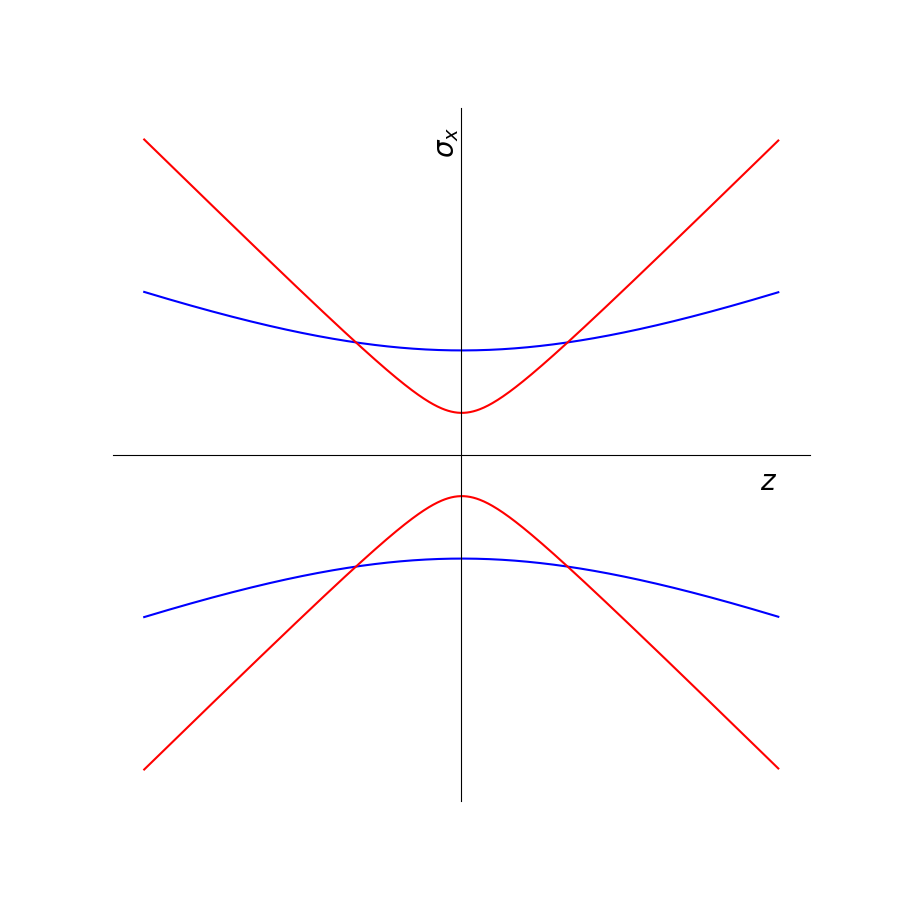}\label{fig-hourglass0}}
    \qquad
    \subfloat[Discretisation of the beam into longitudinal slices of constant beam size.]{\includegraphics[width=0.35\linewidth]{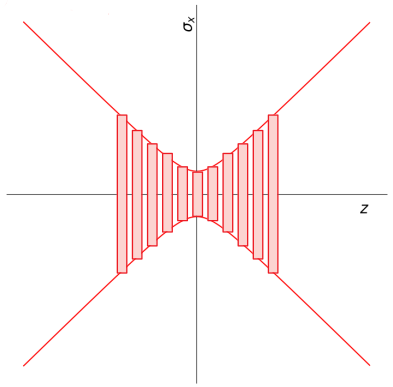}\label{fig-hourglass-slice}}
    \qquad
\end{center}
\caption{Illustration of the hourglass effect and its model based on longitudinal slices.}
\label{fig-hourglass}
\end{figure}
\begin{figure}
\begin{center}
    \subfloat[Start of the collision]{\includegraphics[width=0.3\linewidth]{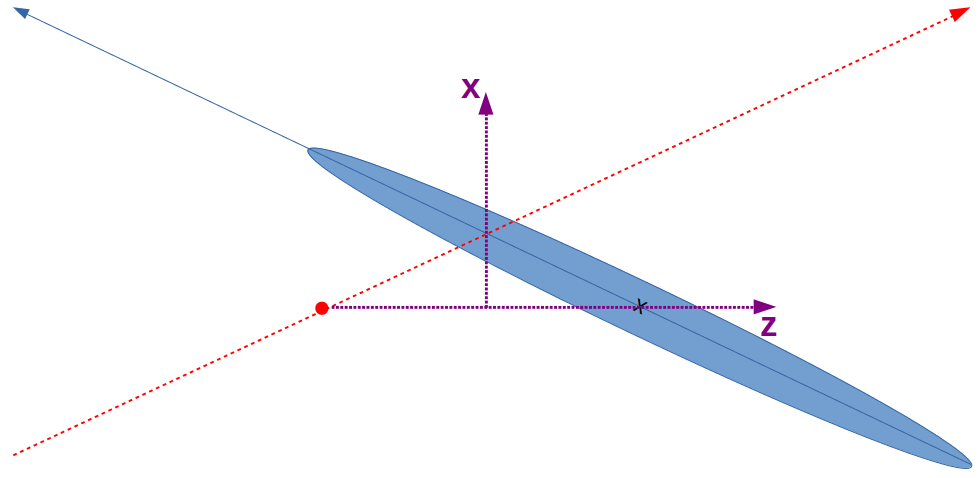}\label{fig-XingStart}}
    \qquad
    \subfloat[Half way through the collision]{\includegraphics[width=0.3\linewidth]{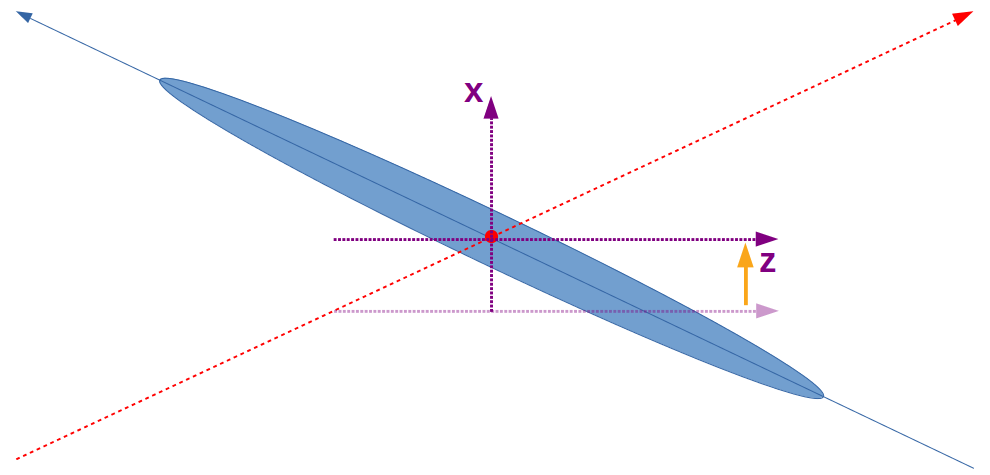}\label{fig-XingMid}}
    \qquad
    \subfloat[End of the collision]{\includegraphics[width=0.3\linewidth]{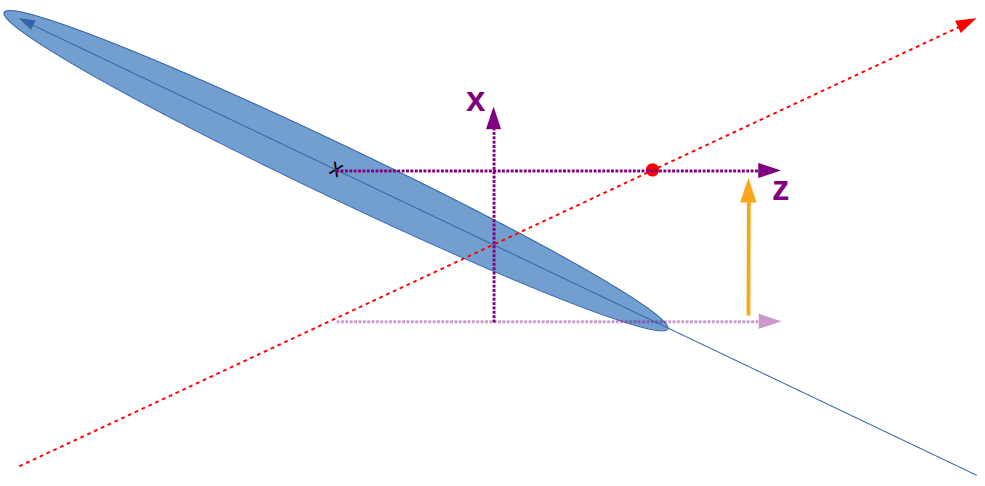}\label{fig-XingEnd}}
    \qquad
    \subfloat[View from boosted frame]{\includegraphics[width=0.3\linewidth]{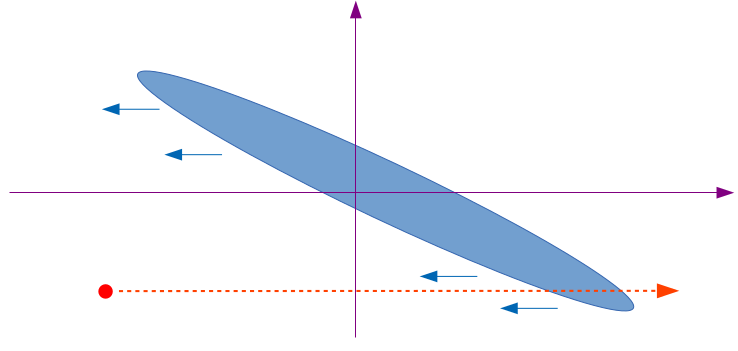}\label{fig-XingBoost}}
    \qquad
    \subfloat[Longitudinal slices in the boosted frame]{\includegraphics[width=0.3\linewidth]{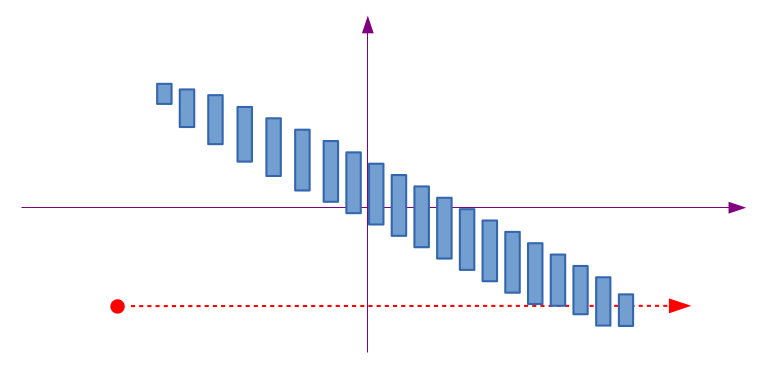}\label{fig-XingSlice}}
    \qquad
\end{center}
\caption{Illustration of the collision of a particle (red dot) with a bunch (blue) featuring a large crossing angle. The~trajectories are marked with red and blue arrows respectively, while the $x$ and $z$ axis of the~co-moving reference frame are marked in purple.}
\label{fig-Xing}
\end{figure}
Above we have considered an infinitely long beam of charge, which is not always representative of realistic configuration. The beams are usually bunched longitudinally and the interaction between the~beams occurs on a finite length. In addition we have negelcted the modification of the particle trajectory while going through the other beam. Nevertheless, these approximations remain valid if:
\begin{itemize}
\item The focusing at the Interaction Point (IP) is weak enough ($\beta^* \ll \sigma_z$, with $\beta^*$ the optical $\beta$ function at the IP and $\sigma_z$ the bunch length).
\item The trajectory of the particle is parallel to the other beam's (Paraxial approximation).
\item The beam-beam force is not strong enough to modify the particle's trajectory over its passage through the beam. (Low disruption, see Section~\ref{sec-linear}.)
\end{itemize}
If one of the approximation is broken, it is convenient to split the beam into small longitudinal slices, such that the approximations are met for the interactions of a particle with each of the slices. For example, in the case of strong focusing ($\beta^* \ll \sigma_z$), the beam size variations at the IP (so-called hourglass effect) can be approximated by a set of longitudinal slices of varying size as is illustrated in Fig.~\ref{fig-hourglass}.

In the case of a crossing angle between the beams, it is convenient to define a reference frame that propagates with the position of the two beams, as illustrated in Figs.~\ref{fig-XingStart},~\ref{fig-XingMid} and~\ref{fig-XingEnd}. In this boosted reference frame, the trajectory of the particle under consideration is co-linear with the movement of the~opposing beam (Fig.~\ref{fig-XingBoost}), such that the paraxial approximation is restored. The collision can now be split into a subset of shorter collisions with longitudinal slices of the opposing beam. At the location of the~collision between the particle and the slice, the beam-beam force is offset transversely due to the~crossing angle (Fig.~\ref{fig-XingSlice}).

Theses so-called bunch length effects introduce a dependence of the beam-beam force on the~longitudinal position of the particles in the opposing beam. In the case of the crossing angle, the forces are transverse in the boosted frame. However once these forces are expressed in the particles' reference frame, the forces acquire a longitudinal component (i.e. an energy change). This enables processes that involve both the transverse and longitudinal motion of the particles, usually referred to as synchrobetatron effects~\cite{SBCPiwinski,SBCHirata,PhysRevAccelBeams.21.031002}.
\section{Orbit and optics distortions}\label{sec-dynamic}
\begin{figure}
\begin{center}
    \subfloat[]{\includegraphics[width=0.45\linewidth]{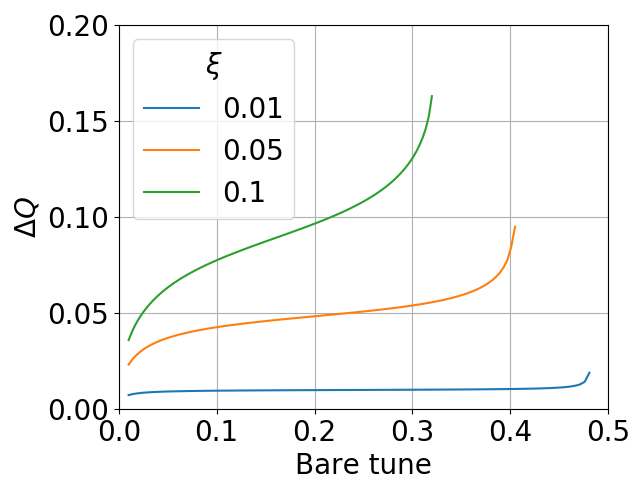}}
    \qquad
    \subfloat[]{\includegraphics[width=0.45\linewidth]{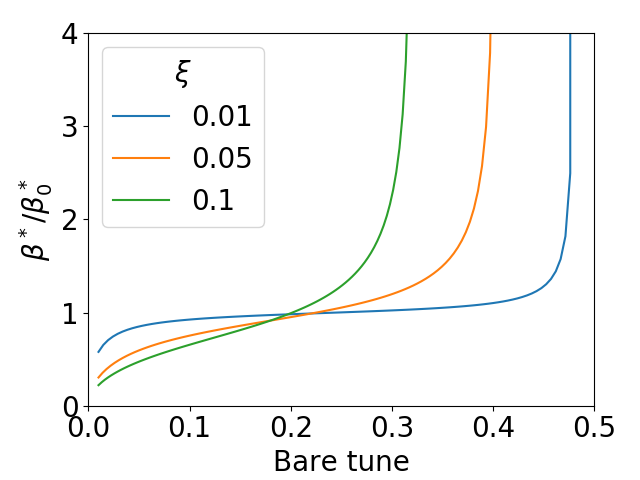}}
    \qquad
\end{center}
\caption{Tune shift and relative change of the optical $\beta$ function at the IP as a function of the machine tune for different beam-beam parameters.}
\label{fig-dynamic}
\end{figure}
A number of beam-beam effects can be modelled by reducing the non-linear beam-beam force to its zero and first order components. When the beams collide without transverse offset, the average of the~force (the coherent force) is zero. However when the beams collider with an offset, the force no longer averages out. The result is a modification of the beam orbit~\cite{PhysRevLett.62.2949}.

As discussed in Section~\ref{sec-bbforce}, by linearising the beam-beam force we treat the beam-beam force as a~quadrupolar error in the lattice and compute the corresponding change of optics and related parameters. Equation~\eqref{eq-bbtuneshift} gives the corresponding impact on the tune. The $\beta$ function is also modified as~\cite{accHandbook}:
\begin{equation}
\frac{\beta^*}{\beta^*_0} = \frac{\sin(2\pi Q_0)}{\sin(2\pi(Q_0+\Delta Q_{BB}))} \label{eq-dynamicbeta}.
\end{equation}
The behaviour of the tune and $\beta$ function at the IP are illustrated in Fig.~\ref{fig-dynamic}. As for quadrupolar errors, the~distortion of the optics by the beam-beam interaction strongly depends on the proximity of the tune to half integer resonances. If the dynamics of the particles is affected by the emission of synchrotron radiation in the lattice, the change of optics caused by the beam-beam interaction will also impact the~radiation integrals and thus modify the equilibrium parameters. These changes are often called dynamic-$\beta$ and dynamic-emittance effects respectively.
\section{Self-consistent solutions}
Up to now we treated the beam-beam force as a static field, just like a dipole or quadrupole error. This is often called the weak-strong regime. The strong beam generates a force that affects the weak beam's property, but the strong beam remains unaffected. This situation was met in proton-anti-proton colliders~\cite{spps,Tevatron}, as the production of an intense anti-proton beam is much more demanding than the one of a proton beam. However most collider feature beams of comparable intensity, such that the distortion of the beam orbit or optics by the beam-beam force will affect both beams' properties and consequently the~beam-beam forces that they experience. In this so-called strong-strong regime, one needs to find the~new configuration of the two beams with the matching beam-beam force. We talk about self-consistent solutions. Such solutions can be relevant for example when considering the orbit of non-uniformly distributed multi-bunch beams~\cite{LEPPACMANOrbit,LHCPACMANOrbit}. Below we rather illustrate the importance of self-consistent solutions with the so-called flip-flop phenomenon.
\subsection{Flip-flop}
The impact of the beam-beam interaction on the $\beta$ function at the IP (Eq.~\eqref{eq-dynamicbeta}) can be extended into a~system of two equations for the properties of the two beam~\cite{chaoFlipFlop}
\begin{equation}
\left\lbrace
\begin{array}{l}
\left(\frac{\beta^*_0}{\beta^*_+}\right)^2 = 1+4\pi\xi_0\cot(2\pi Q_0)\frac{\beta^*_0}{\beta^*_-}-4\pi^2\xi_0^2\left(\frac{\beta^*_0}{\beta_-^*}\right)^2 \\
\left(\frac{\beta^*_0}{\beta^*_-}\right)^2 = 1+4\pi\xi_0\cot(2\pi Q_0)\frac{\beta^*_0}{\beta^*_+}-4\pi^2\xi_0^2\left(\frac{\beta^*_0}{\beta_+^*}\right)^2
\end{array}
\right.
\end{equation}
with $\beta^*_0$ and $\xi_0$ the unperturbed $\beta$ function at the IP and the unperturbed beam-beam parameter respectively. $\beta_\pm$ are the self-consistent solutions of the non-linear system of equations, corresponding to the~$\beta$ function at the IP for the two beams. In most cases, a symmetric solution ($\beta_+=\beta_-$) exists, yet it is different from the one obtained with the weak-strong formula. In specific conditions, there exists additional solution with unequal beam sizes, which break the symmetry of the system. With an external perturbation, it is possible to make the beams switch from the solution with $\beta_+>\beta_-$ to the one with $\beta_+<\beta_-$, thus the name \textit{flip-flop}.
\subsection{Coherent oscillations}\begin{figure}
\begin{center}
    \subfloat[LEP~\cite{bbmodeLEP}]{\includegraphics[width=0.45\linewidth]{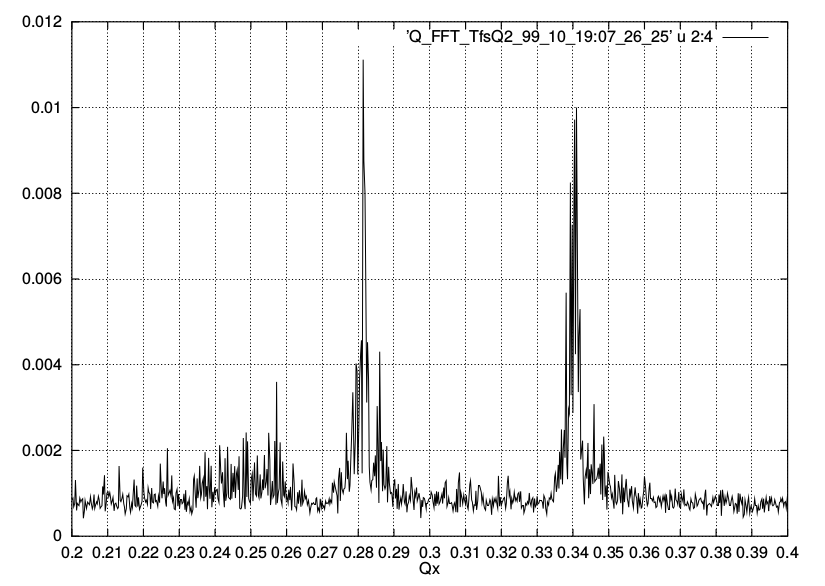}}
    \qquad
    \subfloat[LHC~\cite{bbmodeLHC}]{\includegraphics[width=0.45\linewidth]{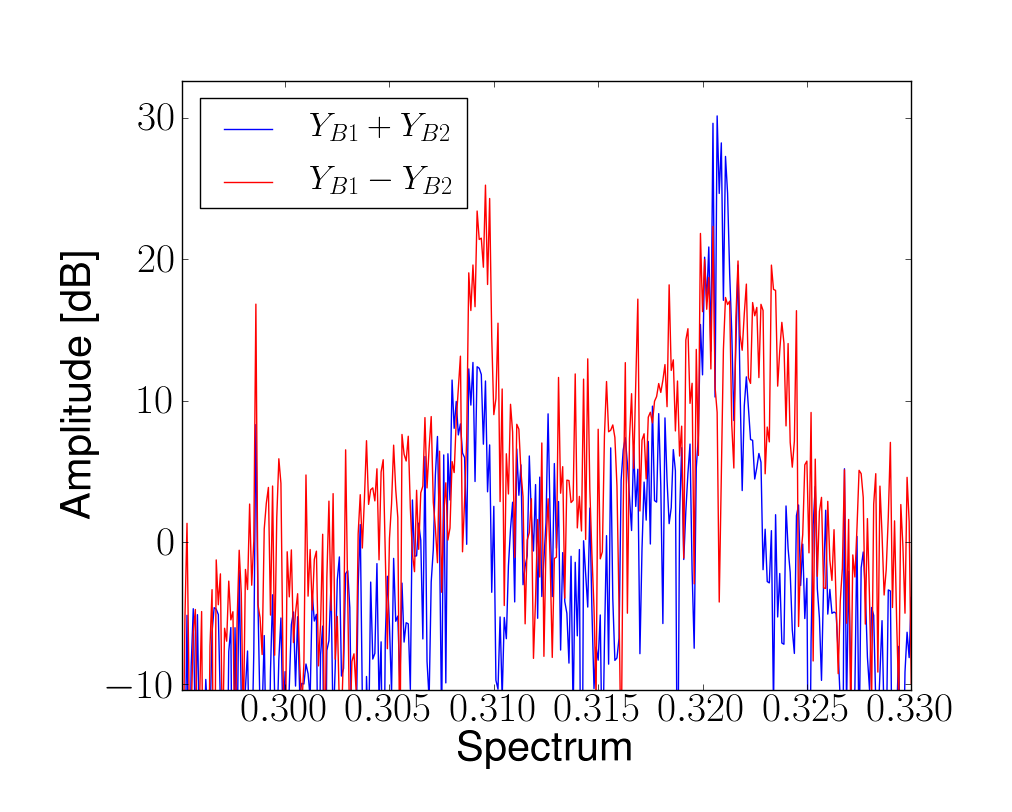}}
    \qquad
\end{center}
\caption{Measurements of beam oscillation spectrum exhibiting a the beam-beam $\sigma$ and $\pi$ modes.}
\label{fig-coherentMode}
\end{figure}
Considering two identical beams colliding in one interaction point, we can write the average the momentum change of the two beams:
\begin{equation}
\left\lbrace
\begin{array}{l}
\Delta p_{x,1} = -K_{BB}(x_2-x_1) \\
\Delta p_{x,2} = -K_{BB}(x_1-x_2),
\end{array}
\right.
\end{equation}
where we have introduced the coherent beam-beam strength $K_{BB}$, which is analogous to the beam-beam strength $k_{BB}$ (Eq.~\eqref{eq-kbb}) but averaged over the beam distribution. It can be obtained using the effective beam-sizes according to the relation given by Eq.~\eqref{eq-effsize}. This so-called rigid bunch model assumes that the~beam distribution is unaffected by the beam-beam force and only the average positions and momentum of the two beams are dynamical variables ($x_1$, $p_{x,1}$, $x_2$, $p_{x,2}$). We can derive two modes of oscillation. If the two beams oscillate in-phase ($\sigma$-mode), the beam-beam force is zero as $x_1=x_2$. The frequency of this mode of oscillation correspond to the bare tune of the machine. If the two beams oscillate out-of-phase ($\pi$-mode), we habe $x_1=-x_2$ and the momentum change due to beam-beam becomes $2K_{BB}$. As for the tune shift (Eq.~\eqref{eq-bbparameter}), we find that the frequency of this mode is shifted by -2$\Xi$, with $\Xi$ the coherent beam-beam tune shift which is again analogous to the beam-beam tune shift $\xi$, but based on the effective beam sizes (Eq.~\eqref{eq-effsize}). Straight forwardly one finds that 2$\Xi=\xi$, thus the $\pi$-mode is shifted from the bare machine tune by the beam-beam parameter $\xi$. Measurements of these modes are shown in Fig.~\ref{fig-coherentMode}. The $\pi$ mode is shifted up or down in the two colliders since they feature opposite and identically charged beams respectively.

While the rigid bunch model gives a good estimates of the frequency of the coherent modes, the non-linearity of the beam-beam force will induce modifications of the beam distribution and thus affect the beam-beam force. With a self-consistent analytical treatment (based on Vlasov equation) it is possible to derive more accurately the frequency shift of the $\pi$-mode~\cite{Yokoya}. This additional impact is well characterised by a form factor (usually called the Yokoya factor), such that the frequency shift of the~$\pi$-mode is given by $Y\xi$, with $Y\approx 1.1-1.3$.
\section{Non-linear effect}
\begin{figure}
\begin{center}
    \subfloat[LHC~\cite{PhysRevAccelBeams.21.081002}]{\includegraphics[width=0.45\linewidth]{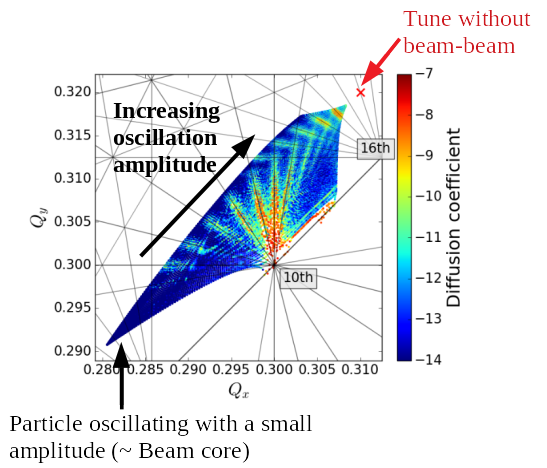}\label{fig-FMA-LHC}}
    \qquad
    \subfloat[DA$\Phi$NE~\cite{PhysRevSTAB.14.014001}]{\includegraphics[width=0.45\linewidth]{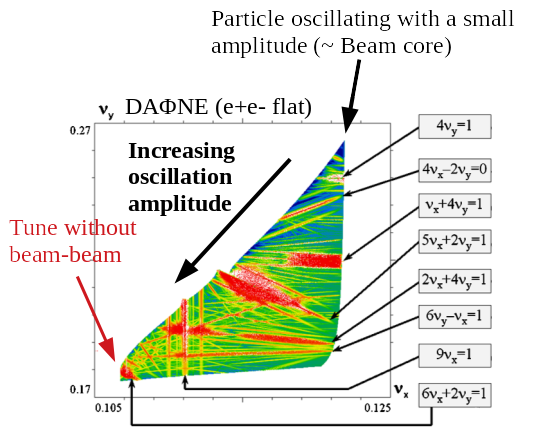}\label{fig-FMA-DAPHNE}}
    \qquad
\end{center}
\caption{Frequency map analysis for a head-on beam-beam interaction in a protron-proton collider (LHC) and an~electron-positron collider (DA$\Phi$NE).}
\label{fig-FMA}
\end{figure}
Due to its strongly non-linear nature, the beam-beam force will cause amplitude detuning as well as drive resonances. The resulting non-linear mechanisms of diffusion may result in a deterioration of the~beam quality through growth of the beam emittances or particle losses. These effects can be visualised using a Frequency Map Analysis~\cite{FMA}, as shown in Fig.~\ref{fig-FMA}. The transverse tunes of particles featuring different initial oscillation amplitudes in the transverse directions $x$ and $y$ are displayed with a color related to their diffusion speed in phase space. We observe that the particles in the beam occupy the space between the bare machine tune and the beam-beam tune shift. The best preservation of the beam quality is obtained when there are no strong non-linear resonance in this space. The maximum beam-beam tune shift is therefore limited by the available tune space without strong resonances. The exact limit is very much dependent on the machine type and design. For example, colliders featuring a strong damping mechanism can tolerate significantly stronger non-linear diffusion mechanism than colliders without damping. Typically in electron-positron colliders the beam-beam tune shift is $\mathcal{O}(10^{-1})$ (see Fig.~\ref{fig-FMA-DAPHNE}), while only ~$\mathcal{O}(10^{-2})$ in hadron colliders (Fig.~\ref{fig-FMA-LHC}).
\section{Beamstrahlung}
In very high energy electron-positron colliders, the deflection of the particles' trajectories by the beam-beam force can lead to sizable emission of synchrotron radiation. The random energy loss experienced by the particles results in bunch lengthening~\cite{beamstrahlungblowup} as well as beam losses when it exceeds the acceptance of the ring. It is characterised by the so-called beamstrahlung lifetime~\cite{PhysRevSTAB.17.041004}. \\
\section{Mitigations}
\subsection{Electron lens}
\begin{figure}
\begin{center}
    \subfloat[Beam-beam forces]{\includegraphics[width=0.4\linewidth]{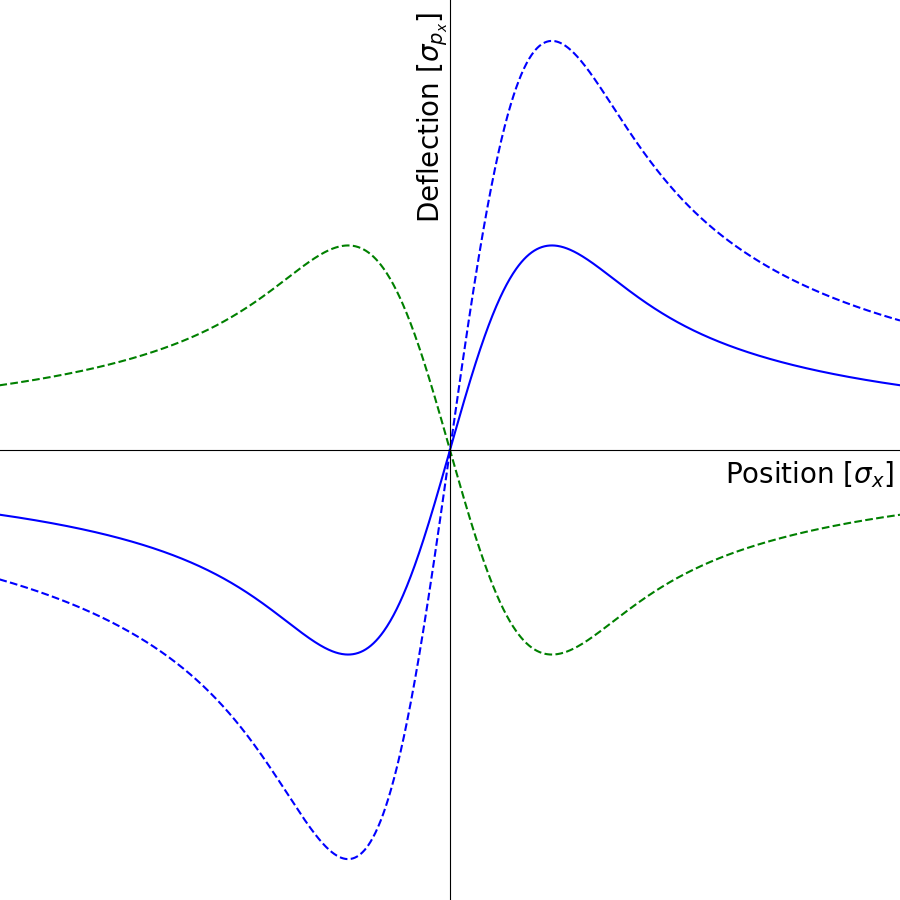}}
    \qquad
    \subfloat[Tune spread measurement~\cite{BBCompRHIC}]{\includegraphics[width=0.45\linewidth]{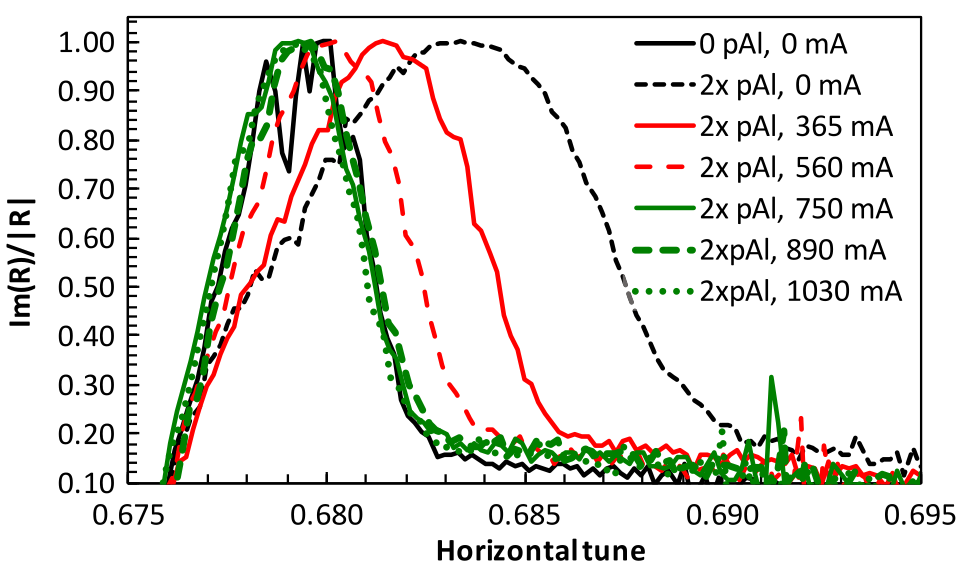}}
    \qquad
\end{center}
\caption{Illustration of the compensation of the total beam-beam force (dasehd blue) by half (dashed green) resulting in half of the beam-beam force (solid blue). The tune spread measured through the beam transfer function shows that the tune spread is reduced by increasing the electron lens current (red and green curves) with respect to the~case without compensation (black curve).}
\label{fig-HOComp}
\end{figure}

In principle, the beam-beam force can be entirely compensated by another beam-beam interaction with a beam of opposite charge. As electron beams are rather easy to produce, beam-beam effects in hadron colliders can efficiently be mitigated by a counter propagating electron beam with the same shape as the other hadron beam, it is called an electron lens. Such a lens was first installed at the Tevatron~\cite{tevatron_lens}, however they were mostly used to correct the bunch-by-bunch tune shifts caused by parasitic long-range beam-beam interactions. The first compensation of non-linear beam-beam effects with an electron lens was achieved at the Relativistic Heavy Ion Collider (RHIC)~\cite{BBCompRHIC}. A reduction of the tune spread (Fig.~\ref{fig-HOComp}) as well as a significant improvement in beam lifetime were obtained.
\subsection{Separation scheme} \label{sec-sep}
\begin{figure}
\begin{center}
    \subfloat[Pretzel scheme at the Sp$\bar{\text{p}}$S]{\includegraphics[width=0.5\linewidth]{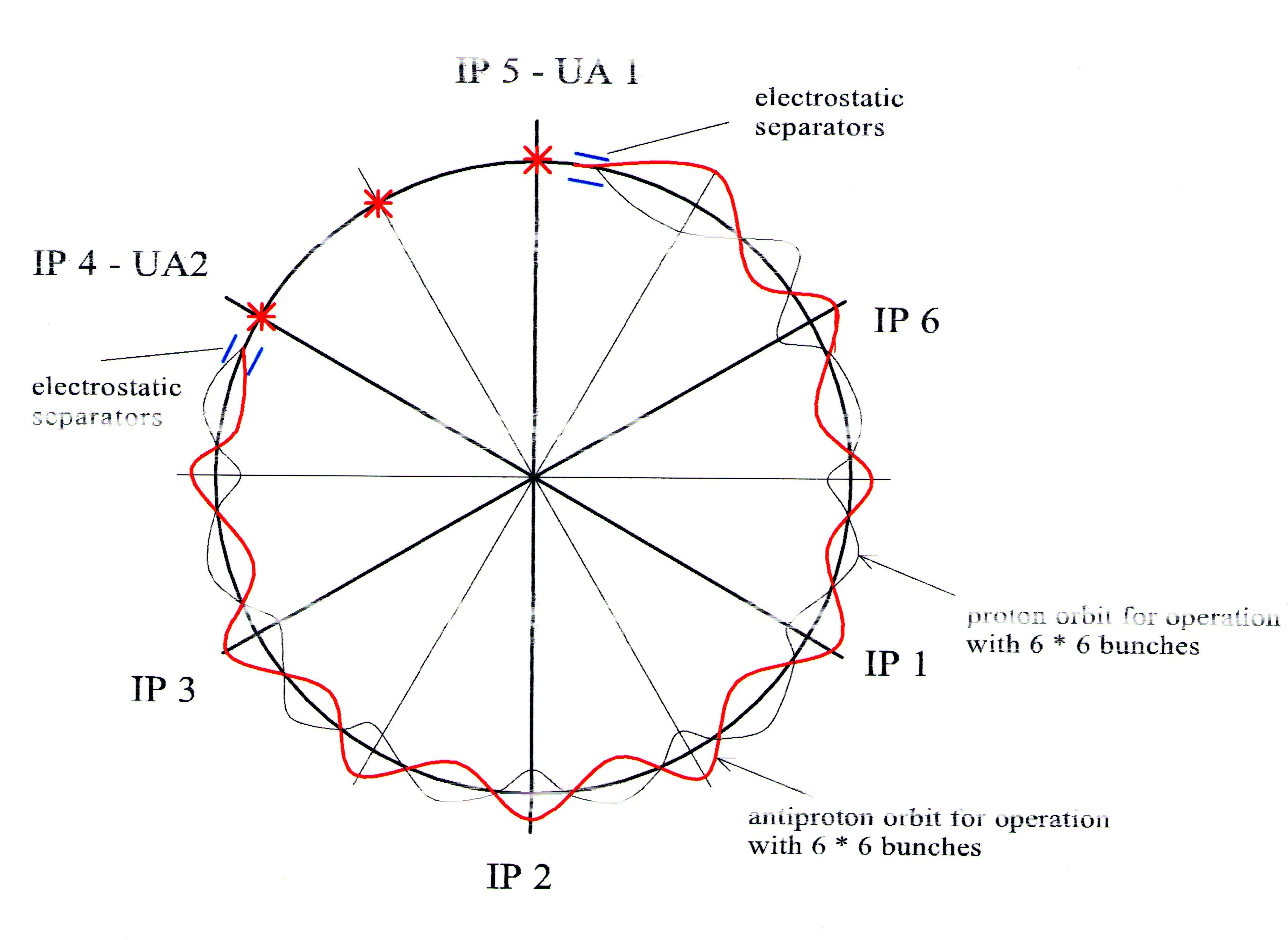}\label{fig-pretzel}}
    \qquad
    \subfloat[Crossing with two beam pipes]{\includegraphics[width=0.4\linewidth]{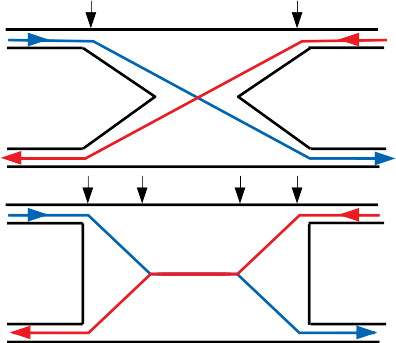}\label{fig-sep_pipe}}
    \qquad
\end{center}
\caption{Illustration of separations schemes for a common beam pipe or two separate ones. In Figure~\ref{fig-sep_pipe}, the black arrows mark the dipole magnet positions in a configuration with a crossing angle at the IP (top) or with full head-on collision at the IP (bottom).}
\label{fig-sep}
\end{figure}
When the strength of the beam-beam forces at the IP is the limiting factor for the performance, modern colliders usually employ beams composed of several bunches. This turned out very challenging in early colliders which profited from the fact that two counter rotating beams with opposite charge can circulate in a single magnetic lattice. Increasing the number of bunches in such a setup where the two beams circulate on the same orbit all along the machine would generate a number of unwanted collisions outside of the detectors. The so-called Pretzel separation scheme was introduced. The idea is to introduce electrostatic plates that generate separate orbits for the two beams yet remaining in a common vacuum chamber all around the machine. The two orbits can be arranged such that the beams collide head-on only at the~center of the desired IP and with a large offset at the other locations (Fig.~\ref{fig-pretzel}). These so-called parasitic interactions become long-range (Fig.~\ref{fig-force-LR}). They do not produce any luminosity and the~associated electromagnetic fields are significantly weaker than the head-on interactions. Nevertheless, when operating with a large number of bunches, there can be numerous of such long-range interactions such that the~performance is eventually limited by their impact on the beam dynamics~\cite{Tevatron,LEP_LR}. To push the~performance further, the two beams must travel in separate beam pipes and share a common chamber only around the IP (Fig.~\ref{fig-sep_pipe}). Most of the parasitic beam-beam interactions can be avoided in this way, yet an important compromise must be made. A short common beam pipe around the IP necessarily implies a~large crossing angle between the beams, leading to a luminosity loss (see Section \ref{sec-CC}). A longer common region allows to accommodate lower crossing angles, yet it implies that some parasitic interactions may take place in the common area. We can compare two different effective approaches, the first one corresponds to the LHC: The common chamber is about a hundred meter long and features few tens of parasitic interactions around each IP. The crossing angle between the beams is limited by the strength of the long-range beam-beam interactions to a few hundreds of $\mu$rad~\cite{LHCDesign} (bottom plot in Fig.~\ref{fig-sep_pipe}). The~second design corresponds to the FCC-ee~\cite{FCCeeDesign}: the parasitic beam-beam interactions are fully prevented by a short common beam pipe (2~m), enforcing a~large crossing angle between the beams of 30~mrad (top plot in Fig.~\ref{fig-sep_pipe}).
\subsubsection{Wire compensation}
In colliders featuring a long common beam pipe, the performance remains limited by the impact of the~parasitic interactions on the beam dynamics. Given that they are long-range, their compensation can be achieved with a current-carrying wire which feature a similar $1/r$ dependence of the field (Fig.~\ref{fig-force-LR}), yet with opposite sign. Dedicated experiments were conducted at the RHIC and the LHC demonstrating the potential of this approach~\cite{PhysRevSTAB.14.091001,LHCWire}.
\subsubsection{Crab cavity} \label{sec-CC}
\begin{figure}
\begin{center}
    \subfloat[Crossing angle]{\includegraphics[width=0.45\linewidth]{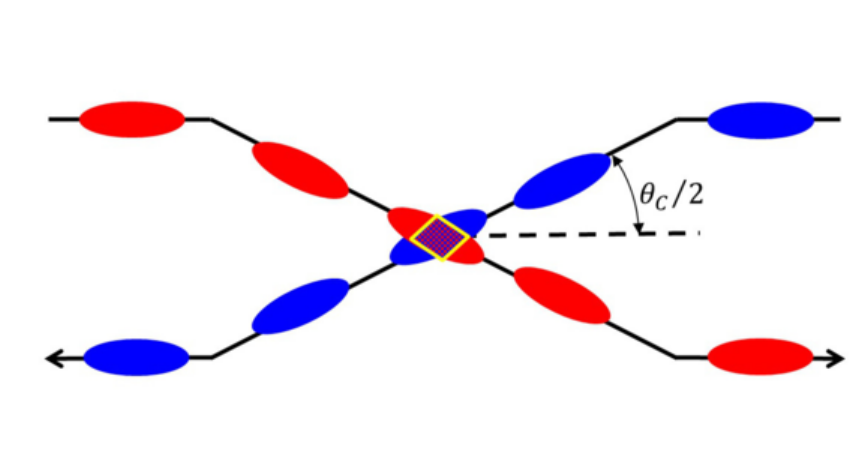}}
    \qquad
    \subfloat[Crab cavities]{\includegraphics[width=0.45\linewidth]{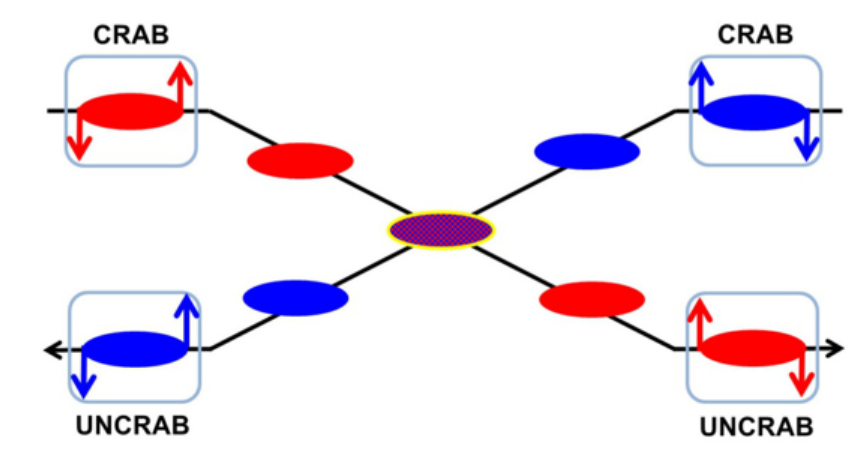}}
    \qquad
\end{center}
\caption{Beams colliding with a crossing angle, with or without crab cavities.}
\label{fig-crabXing}
\end{figure}
Rather than compensating the force itself, it is also possible to alleviate the detrimental impact of the~crossing angle on the luminosity by tilting the bunch at the interaction point, thus avoiding the~geometric loss of luminosity (Fig.~\ref{fig-crabXing}). This can be achieved by means of so-called crab cavities. They feature a time-dependent transverse deflecting field such that the head and the tail of the bunch are kicked in opposite directions. The KEKB was operated with crab cavities~\cite{KEKBCavities}. They are part of the design of the~HL-LHC~\cite{HLLHCDesign}, FCC-hh~\cite{FCChhDesign} and the Electron-Ion Collider (EIC)~\cite{EIC}.
\subsubsection{Crab waist}
\begin{figure}
\begin{center}
    \subfloat[Without crab sextupoles]{\includegraphics[width=0.45\linewidth]{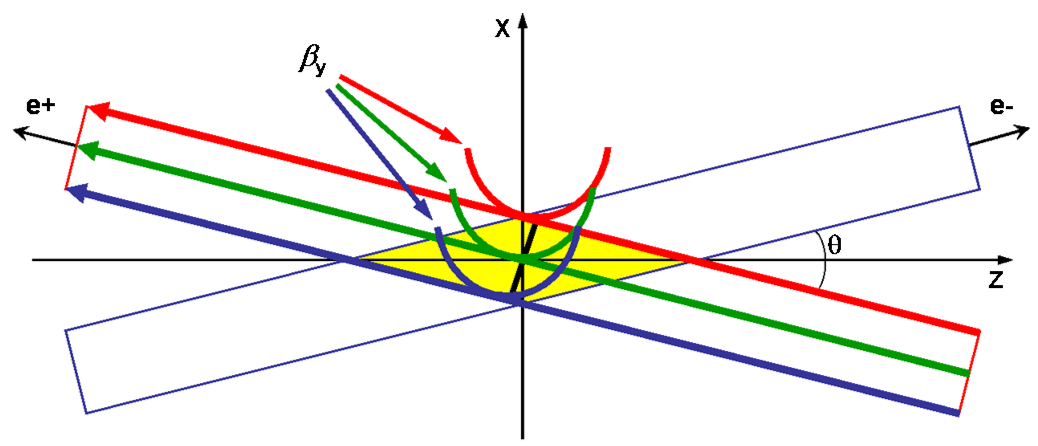}\label{fig-traj_noCW}}
    \qquad
    \subfloat[With crab sextupoles]{\includegraphics[width=0.45\linewidth]{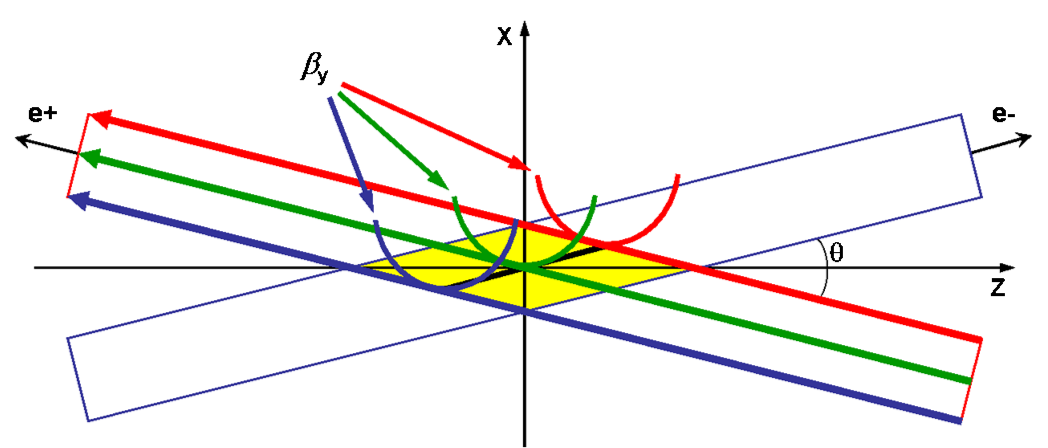}\label{fig-traj_CW}}
    \qquad
    \subfloat[Without crab sextupoles]{\includegraphics[width=0.48\linewidth]{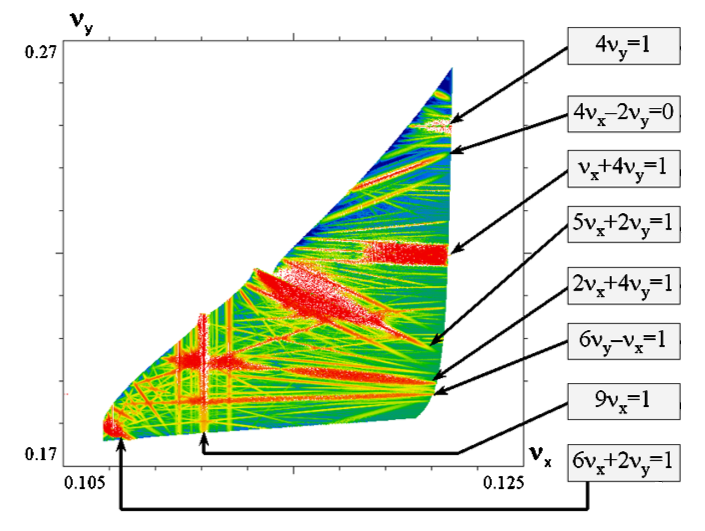}\label{fig-FMA_noCW}}
    \qquad
    \subfloat[With crab sextupoles]{\includegraphics[width=0.42\linewidth]{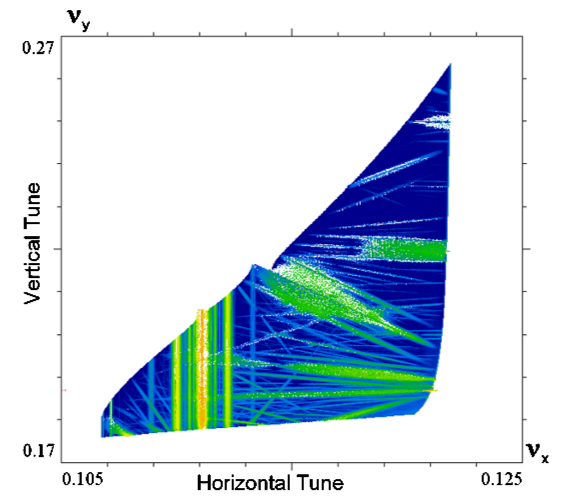}\label{fig-FMA_CW}}
    \qquad
\end{center}
\caption{The trajectories and $\beta$ functions of particles with different transverse offsets in the plane of the~crossing angle are shown in red, green and blue with and without crab sextupoles (top plots). The~other beam is marked with a blue box. The sextupoles shifts the focal point (minimum of the $\beta$ function) to align it with the other beam trajectory, thus reducing the modulation of the beam-beam force when the~particles perform betatron motion. The~corresponding frequency map analysis are shown in the~lower plots. (Courtesy \cite{PhysRevSTAB.14.014001})}
\label{fig-crabWaist}
\end{figure}
An alternative to the crab crossing scheme is the so-called crab waist scheme. While they feature confusable names, they are very different in nature. While in the crab crossing, the geometric loss of luminosity is recovered with RF cavities, in the crab waist scheme the luminosity is increased by further pushing the~beam parameters (e.g. the bunch intensity), thus leading to stronger beam-beam forces at the interaction point. However, the diffusion mechanisms enabled by these strong non-linear forces are mitigated by the~introduction of two sextupoles, one on each side of the IP. The role of these sextupole is to shift the focal point of particles with different transverse offsets in the crossing angle plane (Figs. \ref{fig-traj_noCW} and \ref{fig-traj_CW}) such that it follows the orbit of the other beam. This shift reduces the modulation of the beam-beam force and thus mitigates several resonances. The beneficial impact on the beam dynamics is visible in Figs.~\ref{fig-FMA_noCW} and~\ref{fig-FMA_CW}. DA$\Phi$NE operated with this scheme, yielding a large improvement of the luminosity~\cite{PhysRevLett.104.174801}. Currently SuperKEKB operates with the crab waist scheme~\cite{SuperKEKB}. This scheme is also part of the~design of the FCC-ee~\cite{FCCeeDesign}.
\section{Linear colliders} \label{sec-linear}
\begin{figure}
\begin{center}
\subfloat[Trajectories featuring strong distortion]{\includegraphics[width=0.45\linewidth]{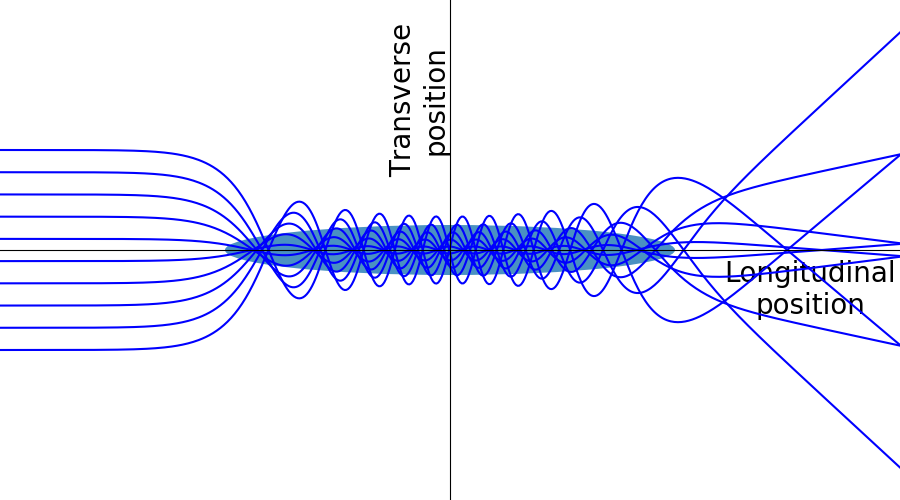}\label{fig-disruption}}
\qquad
\subfloat[Luminosity spectrum~\cite{schulteCAS}]{\includegraphics[width=0.45\linewidth]{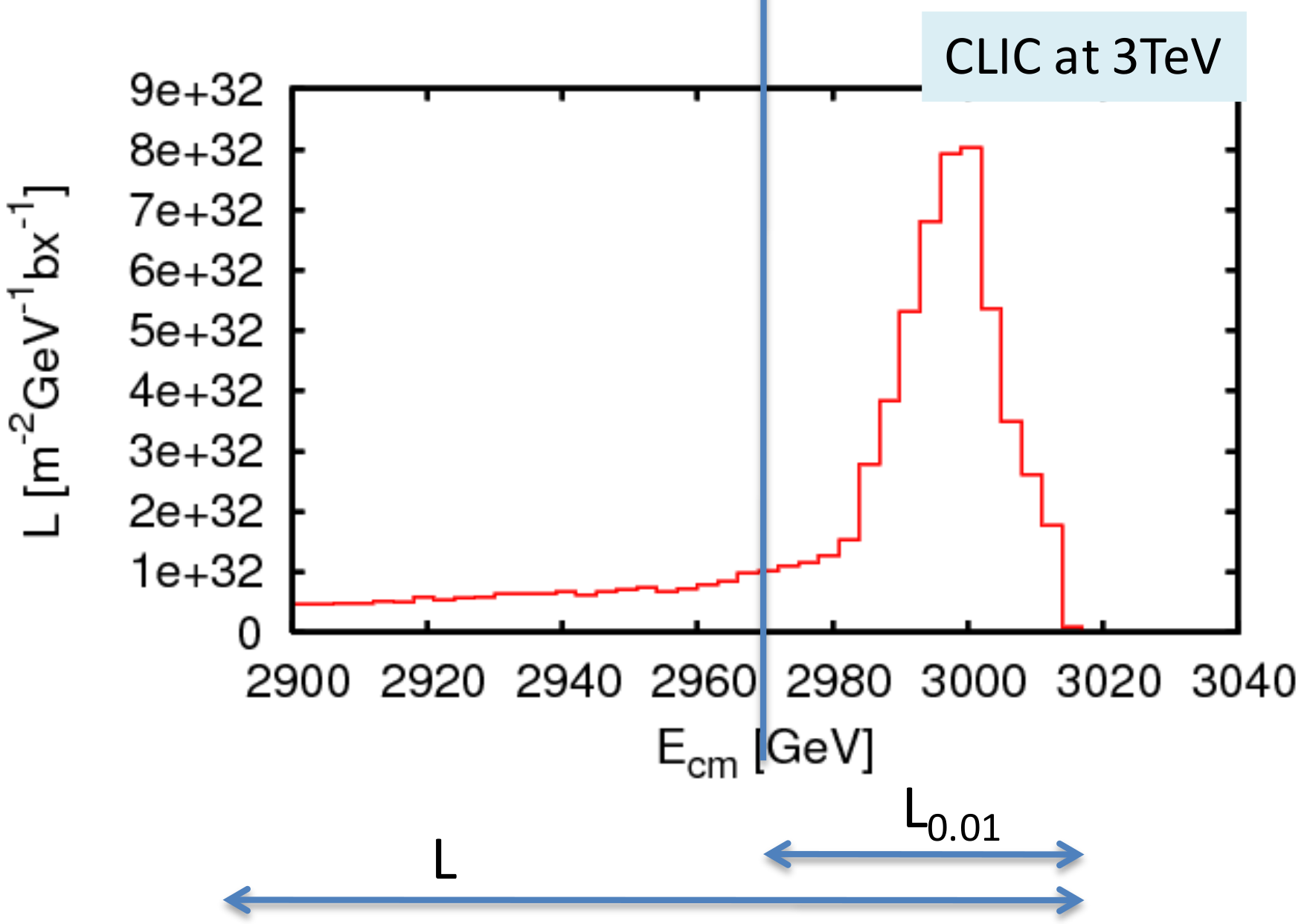}\label{fig-beamstrahlungCLIC}}
\qquad
\end{center}
\caption{Illustration of beam-beam effects in linear colliders. In (b), we observe that most of the collision are spread around the design energy of 3~TeV which stems from the natural energy spread of the beam at the output of the~accelerating section, while the long tail at lower energies is caused by beamstrahlung.}
\label{fig-lin}
\end{figure}
The dynamics of beam-beam interactions in linear colliders differs significantly from the ones in circular colliders. The fact that the particle trajectories do not have to remain stable over many turns allows for significantly stronger beam-beam forces. While in circular colliders the particle trajectories are slightly deflected during their passage thought the other beam, in linear colliders the motion of the particles is fully disrupted by the strong focusing force generated by the other beam. The motion is illustrated in Fig.~\ref{fig-disruption}, it is well characterised by the so-called disruption parameter, corresponding to the ratio of the~bunch length to the beam-beam focal length (i.e. $1/k_{bb}$)~\cite{schulteCAS}:
\begin{equation}
    D_i = \frac{2Nr_0\sigma_z}{\gamma\sigma_i(\sigma_x+\sigma_z)}
\end{equation}
with $i=x$ or $y$. If the distortion is high, the beam distribution evolution during the interaction becomes quite complex and is usually described by means of numerical simulations~\cite{schulteCAS}. Through its additional strong focusing force, the beam-beam interaction tends to further increase the luminosity with respect to the one that would be expected based on formulas that neglect the distortion. The main limitation that arise from the beam-beam interaction is the generation of strong beamstrahlung, such that particles may lose a significant part of their energy before they collide. As a result the center of mass energy of these collisions is lower than the design value. This reduces the 'useful' luminosity, i.e. the amount of collisions close to the desired energy, and increases the spread of the collision energies, as illustrated in Fig. \ref{fig-beamstrahlungCLIC}. Since the knowledge of the collision energy is key for precision measurements, the spread can become a limitation. 

\bibliographystyle{IEEEtran}
\bibliography{CAS_BeamBeam}

\end{document}